\documentclass[aps,prl,superscriptaddress,amsmath,amssymb,twocolumn,10pt,floatfix]{revtex4-2}
\usepackage{graphicx}
\usepackage{dcolumn}
\usepackage{bm}

\begin{document}
\title{Retrieving Metrological Information from Noisy Squeezed Spin States via Bayesian Optimization}

\author{Enze Li}
\thanks{These three authors contributed equally to this work.}
\affiliation{Department of Computer and Network Engineering, The University of Electro-Communications, 1-5-1 Chofugaoka, Chofu, Tokyo 182-8585, Japan}
\author{Yunjie Cai}
\thanks{These three authors contributed equally to this work.}
\affiliation{Department of Information and Electronic Engineering, Faculty of Science and Engineering, Teikyo University, 1-1 Toyosodai, Utsunomiya-shi, Tochigi 320-8551, Japan}
\author{Jingwei Zhang}
\thanks{These three authors contributed equally to this work.}
\affiliation{Department of Information and Electronic Engineering, Faculty of Science and Engineering, Teikyo University, 1-1 Toyosodai, Utsunomiya-shi, Tochigi 320-8551, Japan}
\author{Emi Yukawa}
\email{emi.yukawa@cat.phys.s.u-tokyo.ac.jp}
\affiliation{Department of Physics, The University of Tokyo, 7-3-1 Hongo, Bunkyo-ku, Tokyo 113-0033, Japan}
\date{\today}

\begin{abstract}
Frequency inhomogeneity can obscure the metrological advantage of squeezed spin states, making the choice of readout crucial.
We study interacting spin-$1/2$ ensembles with static local frequency disorder and optimize a common collective readout over independent disorder realizations.
Using Bayesian optimization of the full measurement-outcome distribution, we identify observables combining dipole and quadrupole operators that yield greater classical Fisher information than the optimized dipole readout considered here.
For ensembles of up to 17 spins, their disorder-averaged CFI surpasses the standard quantum limit at both disorder strengths studied, including a regime where the dipole readout remains below it.
For the larger ensembles studied, fixing the encoding axis to the direction that maximizes the disorder-averaged quantum Fisher information retains nearly the same performance.
These results identify a route to extracting metrological information from frequency-disordered spin ensembles through the choice of collective observable.
\end{abstract}
\maketitle

\emph{Introduction.---}
Entanglement can improve measurement sensitivity beyond the standard quantum limit (SQL).
Squeezed spin states provide a well-established route to this enhancement in Ramsey interferometry~\cite{PhysRevA.47.5138,PhysRevA.46.R6797,RevModPhys.90.035005}, with realizations in diverse physical platforms~\cite{doi:10.1073/pnas.0901550106,Gross:2010aa,Riedel:2010aa,PhysRevLett.114.043604,doi:10.1126/science.aad9958,Bornet:2023aa,Eckner:2023aa,Franke:2023aa,Miller:2024aa}.
Recent experiments have demonstrated squeezed spin states in nitrogen-vacancy ensembles~\cite{Wu:2025aa} and strongly interacting solid-state spin platforms~\cite{6z22-4z36}, including cavity-mediated one-axis-twisting dynamics in rare-earth-ion ensembles~\cite{fukumori2026cavitymediatedcoherenceprotectiononeaxis}.
Inhomogeneous spin frequencies present an important challenge for collectively enhanced metrology in these systems~\cite{Dooley_2016,fukumori2026cavitymediatedcoherenceprotectiononeaxis}.

The loss of conventional spin squeezing need not imply the loss of all metrological usefulness.
Fisher-information measurements have revealed gains beyond those captured by the Wineland squeezing parameter~\cite{doi:10.1126/science.1250147}.
Gessner, Smerzi, and Pezz\`e developed a hierarchy of nonlinear squeezing parameters by optimizing moment-based sensitivities over accessible collective spin observables up to second order~\cite{PhysRevLett.122.090503}.
Related work established sensitivity scaling for nonlinear readouts of one-axis-twisted states, including collective dephasing and particle loss~\cite{PhysRevLett.127.160501,CRPHYS_2022__23_G1_1_0}, and experimentally characterized nonlinear squeezing in superconducting qubits~\cite{PhysRevLett.128.150501}.
These developments motivate a complementary question: how much information can a common low-order collective readout extract from states prepared with independent local frequency offsets?
Unlike a fluctuating collective field, such offsets generally break conservation of total spin during state preparation.

Here we address this question by maximizing the disorder-averaged classical Fisher information (CFI) of projective measurements of a chosen collective dipole--quadrupole observable.
We use the full outcome probabilities, rather than only their first two moments, and optimize both the observable and a common pre-encoding rotation using Bayesian optimization (BO).
For $N=5$--17 spins and two disorder strengths, the achieved rank-2 readouts outperform the rank-1 protocol and reach sub-SQL sensitivity.
We further identify when a QFI-guided choice of encoding axis can simplify the search without appreciably reducing its performance.

\emph{Disordered state preparation.---}
We consider $N$ spin-$1/2$ particles with dimensionless collective operators $\hat S_\mu=\sum_i\hat s_\mu^{(i)}$, where $\hat s_\mu^{(i)}=\hat\sigma_\mu^{(i)}/2$ is the spin operator of particle $i$, $\hat\sigma_\mu^{(i)}$ is the corresponding Pauli matrix, and $\mu=x,y,z$.
In a frame rotating at the center of the frequency distribution, the Hamiltonian is
\begin{equation}
 \hat H=\hbar\chi\left[-\hat S_+\hat S_-+
 \sum_{i=1}^N\delta\tilde\omega^{(i)}\hat s_z^{(i)}\right],
 \label{eq:Hamiltonian}
\end{equation}
where $\hat S_\pm=\hat S_x\pm i\hat S_y$ and $\chi>0$ is the uniform all-to-all interaction strength.
The static dimensionless offsets are drawn independently from the normal distribution $\mathcal N(0,r^2)$, with mean zero and standard deviation $r$; the physical frequency width is therefore $\sigma_\omega=\chi r$.
Such an effective interaction can arise by adiabatically eliminating a far-detuned bosonic mode~\cite{PhysRevA.56.2249,PhysRevA.85.032111,Dooley_2016}.
To focus exclusively on static frequency inhomogeneity, we neglect dissipative processes and retain unitary evolution for each disorder realization.

\begin{figure}[tb]
 \includegraphics[width=\columnwidth, bb = 0 0 244.8 165.6]{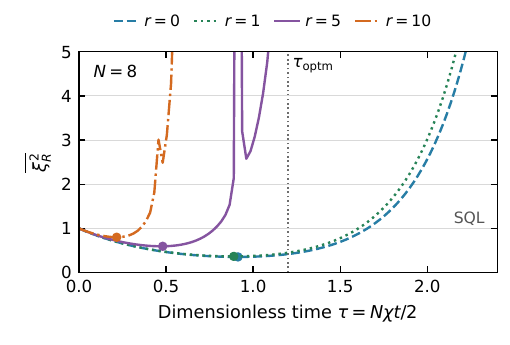}
 \caption{\label{fig:sqparam}Arithmetic mean of the realization-wise Wineland squeezing parameter over 625 disorder samples for $N=8$ at $r=0,1,5,10$, plotted against $\tau=N\chi t/2$.
 Dots mark the minima of the displayed curves.
 Finite-size and disorder effects shift these minima away from the ideal asymptotic reference time in Eq.~(\ref{eq:time}).
 Increasing disorder raises the minima and narrows the interval with $\overline{\xi_R^2}<1$; values above 5 lie outside the displayed range.}
\end{figure}

Starting from the $x$-polarized coherent spin state (CSS)~\cite{PhysRevA.6.2211},
$|\Psi_{\rm init}\rangle=\bigotimes_i[(|\uparrow\rangle_i+|\downarrow\rangle_i)/\sqrt{2}]$,
we prepare $|\psi_k\rangle=e^{-i\hat H_k t_{\rm optm}/\hbar}|\Psi_{\rm init}\rangle$ for realization $k$.
At $r=0$, $-\hat S_+\hat S_-=-\hat{\bm S}^{2}+\hat S_z^2-\hat S_z$ reduces, within the $S=N/2$ sector, to one-axis twisting plus a uniform $z$ rotation.
We use the ideal large-$N$ squeezing time~\cite{PhysRevA.47.5138,Yukawa_2016}
\begin{equation}
 \begin{aligned}
 t_{\rm optm}&=\frac{3^{1/6}}{\chi N^{2/3}},\\
 \tau_{\rm optm}&=\frac{N\chi t_{\rm optm}}{2}
 =\frac{1}{2}(\sqrt{3}N)^{1/3}.
 \end{aligned}
 \label{eq:time}
\end{equation}
This is a common reference time, not a time reoptimized for each finite $N$ or disorder strength.
The Wineland squeezing parameter is
\begin{equation}
 \xi_R^2=\frac{N\min_{\bm n\perp\langle\hat{\bm S}\rangle}
 \langle(\Delta\hat S_{\bm n})^2\rangle}
 {|\langle\hat{\bm S}\rangle|^2},
 \label{eq:squeezing}
\end{equation}
with $\hat S_{\bm n}=\bm n\cdot\hat{\bm S}$.
For ideal one-axis twisting, $\xi_R^2\simeq(3^{2/3}/2)N^{-2/3}$ at this time in the large-$N$ limit~\cite{PhysRevA.47.5138,Yukawa_2016}.
Figure~\ref{fig:sqparam} illustrates how increasing frequency disorder degrades conventional squeezing; at the reference time for $N=8$, the mean $\xi_R^2$ exceeds unity for both $r=5$ and $10$.

\emph{Disorder-averaged Fisher information.---}
After preparation, the same collective spin rotation $\hat R(\bm n,\eta)=e^{-i\eta\bm n\cdot\hat{\bm S}}$, corresponding to an SO(3) rotation, is applied to every realization, followed by ideal phase encoding $e^{-i\theta\hat S_z}$.
For magnetometry, $\theta=\gamma B t_{\rm msr}$, where $\gamma$ is the gyromagnetic ratio and $t_{\rm msr}$ the sensing time.
We treat this encoding separately from the disordered preparation dynamics.
The QFI bounds the standard deviation of a locally unbiased phase estimator through $\Delta\theta\geq(\nu F_Q)^{-1/2}$ for $\nu$ independent repetitions~\cite{helstrom1969quantum,holevo2011probabilistic,PhysRevLett.72.3439,BRAUNSTEIN1996135}.
The CSS benchmark is $F_Q=N$.

For each $(N,r)$, we generate $N_{\mathrm{s}}=\lfloor5000/N\rfloor$ independent disorder realizations.
The number 5000 keeps the numerical sampling budget $N_{\mathrm{s}}N$ approximately constant across sizes; $N$ is the number of mutually interacting spins in each calculation.
We average realization-wise Fisher informations with equal weights, retaining the realization labels while keeping the controls common.
These averages describe performance conditional on fixed disorder realizations, not the Fisher information of the mixed state obtained by forgetting their identity.
This averaging assumes that each static realization, or equivalently its calibrated conditional outcome distribution, remains identifiable.
If the offsets instead fluctuate unpredictably between shots and the realized offsets are not recorded, the probabilities must be averaged first as $p(a|\theta)=N_{\mathrm{s}}^{-1}\sum_{k=1}^{N_{\mathrm{s}}}p_k(a|\theta)$ before calculating the CFI, which defines a different estimation problem.

\begin{figure}[tb]
 \includegraphics[width=\columnwidth, bb = 0 0 244.8 259.2]{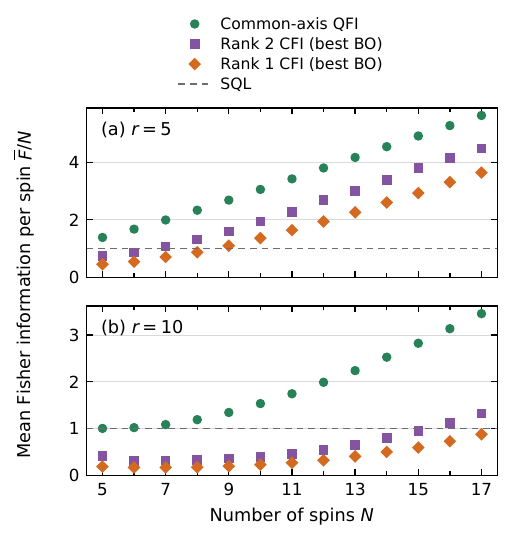}
 \caption{\label{fig:fisher}Common-axis QFI and achieved CFI per spin for (a) $r=5$ and (b) $r=10$ at the preparation time in Eq.~(\ref{eq:time}).
 The CFI markers show the best of 20 full-rotation BO runs for each readout family.
 The dashed line is the SQL, $F/N=1$.
 Rank-2 readout reaches the SQL crossing at $N=7$ for $r=5$ and $N=16$ for $r=10$; the rank-1 reference crosses at $N=9$ for $r=5$ and remains below the SQL through $N=17$ for $r=10$.
 Markers denote the calculated data.}
\end{figure}

For the pure prepared states, the largest mean QFI attainable with one common rotation is
\begin{align}
 \bar F_Q&=\max_{\hat R}\frac{1}{N_{\mathrm{s}}}\sum_{k=1}^{N_{\mathrm{s}}}
 4\operatorname{Var}_{\psi_k}(\hat R^\dagger\hat S_z\hat R)
 =4\lambda_{\max}(\bar C),\label{eq:QFI}\\
 \bar C_{\mu\nu}&=\frac{1}{N_{\mathrm{s}}}\sum_{k=1}^{N_{\mathrm{s}}}\left[
 \frac{\langle\hat S_\mu\hat S_\nu+\hat S_\nu\hat S_\mu\rangle_k}{2}
 -\langle\hat S_\mu\rangle_k\langle\hat S_\nu\rangle_k\right],\nonumber
\end{align}
with $\mu,\nu=x,y,z$ and $k$ labeling the disorder realization.
An optimal rotation satisfies $\hat R^\dagger\hat S_z\hat R=\bm u_{\max}\cdot\hat{\bm S}$, where $\bm u_{\max}$ is the principal eigenvector of $\bar C$.
This common-axis QFI is an upper bound on our readout performance; we do not assume that a measurement restricted to collective observables can saturate this bound.
It differs from averaging QFIs optimized independently for each realization.
Figure~\ref{fig:fisher} shows that $\bar F_Q/N$ exceeds unity and increases over the range studied for both $r=5$ and $10$.

\emph{Optimizing a collective readout.---}
Our rank-2 search uses the restricted five-component family
\begin{equation}
 \begin{aligned}
 \hat A&=c_1\hat S_y+c_2\hat D_{xy}+c_3\hat Q_{xy}\\
       &\quad+c_4\hat Q_{yz}+c_5\hat Q_{zx},\qquad
 \sum_{i=1}^5 c_i^2=1,
 \end{aligned}
 \label{eq:observable}
\end{equation}
with real coefficients and
\begin{align}
 \hat D_{xy}&=\frac{\hat S_x^2-\hat S_y^2}{\sqrt{N-1}},&
 \hat Q_{\mu\nu}&=\frac{\hat S_\mu\hat S_\nu+\hat S_\nu\hat S_\mu}{\sqrt{N-1}}.
 \label{eq:quadrupoles}
\end{align}
The normalization gives these four quadrupoles the same Hilbert--Schmidt norm as $\hat S_y$ on the full $2^N$-dimensional space.
Equation~(\ref{eq:observable}) defines the chosen search family rather than the most general combination of all dipole and quadrupole components; our BO results therefore do not establish the optimum over all rank-1 and rank-2 observables.
We assume projective measurement onto the eigenspaces $\hat P_a$ of $\hat A$ associated with its eigenvalues $a$, with probabilities
\begin{equation}
 p_k(a|\theta)=\langle\psi_k|\hat R^\dagger e^{i\theta\hat S_z}
 \hat P_a e^{-i\theta\hat S_z}\hat R|\psi_k\rangle.
 \label{eq:probability}
\end{equation}
The objective is the mean CFI at the local operating point $\theta=0$,
\begin{equation}
 \bar F_C(\hat R,\hat A)=\frac{1}{N_{\mathrm{s}}}\sum_{k=1}^{N_{\mathrm{s}}}F_{C,k},\quad
 F_{C,k}=\left.\sum_{a:p_k>0}\frac{(\partial_\theta p_k)^2}{p_k}\right|_{\theta=0}.
 \label{eq:CFI}
\end{equation}
We use analytic probability derivatives and omit numerical terms with $p_k\leq10^{-12}$.
For $N_{\mathrm{s}}$ independently interrogated, labeled ensembles, the total CFI is $N_{\mathrm{s}}\bar F_C$; thus $\bar F_C>N$ beats the CSS benchmark with the same total spin resource $N_{\mathrm{s}}N$.
Unlike the moment-based sensitivity optimized in nonlinear squeezing theory~\cite{PhysRevLett.122.090503}, this objective uses the entire spectral outcome distribution.
For a fixed state, encoding, and observable, the corresponding hierarchy is
\begin{equation}
 \frac{|\partial_\theta\langle\hat A\rangle|^2}
 {\langle(\Delta\hat A)^2\rangle}\leq F_C[\hat A]\leq F_Q.
 \label{eq:hierarchy}
\end{equation}
Thus, our full-distribution objective differs from the moment-based objective of Ref.~\cite{PhysRevLett.122.090503}, while our five-component family is contained in the larger second-order operator space considered there.

We optimize three rotation variables and five coefficient variables, normalizing the latter at each evaluation.
The rank-1 reference optimizes the same pre-encoding rotation with the readout fixed to $\hat S_y$.
BO is useful for costly quantum-control objectives~\cite{PRXQuantum.1.020322}; here it receives probabilities calculated from the model, without measurement shot noise.
Using the software in Ref.~\cite{bo}, we perform 20 runs per case: five random seeds and expected-improvement parameters $0.01,0.03,0.05,0.1$.
Each full rank-2 run uses 80 initial evaluations and 400 BO iterations; rank-1 runs use 40 and 400, respectively.
All total-spin sectors and their multiplicities are retained in evaluating the collective spectral probabilities, including degeneracies shared between sectors.
We report the best values found, denoted $\bar F_C^{\rm BO}$, rather than certified global maxima.

\emph{Recovered information.---}
For every size in Fig.~\ref{fig:fisher}, the achieved rank-2 CFI exceeds its rank-1 counterpart.
At $r=5$, including quadrupoles advances the first observed SQL crossing from $N=9$ to $N=7$.
At $r=10$, the rank-2 readout crosses at $N=16$, whereas the rank-1 reference remains below the SQL up to $N=17$.
At $N=17$, the achieved rank-2 (rank-1) values are $\bar F_C^{\rm BO}=76.14$ ($61.84$) for $r=5$ and $\bar F_C^{\rm BO}=22.51$ ($14.89$) for $r=10$.
Thus stronger disorder delays, but does not eliminate, the recovered gain within this range.
These comparisons concern the specified protocols and finite BO budgets, not a proof excluding every possible linear readout.

\begin{figure}[tb]
 \includegraphics[width=\columnwidth, bb =  0 0 244.8 151.2]{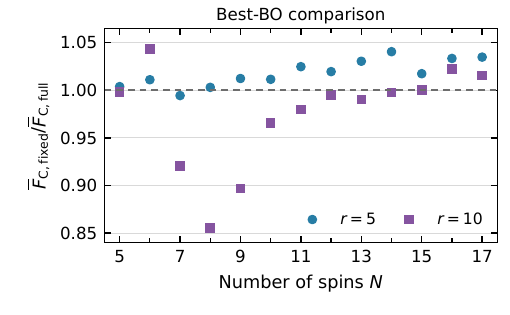}
 \caption{\label{fig:axis}Ratio of the best mean CFI $\bar F_C^{\rm BO}$ found with the QFI-optimal encoding axis fixed to that found with the full three-parameter rotation.
 Both searches use Eq.~(\ref{eq:observable}) and the same disorder realizations.
 Ratios above one are possible for finite BO searches, although the exact constrained optimum cannot exceed the unrestricted one.}
\end{figure}

To test whether the QFI geometry can guide the search, we constrain $\hat R^\dagger\hat S_z\hat R=\bm u_{\max}\cdot\hat{\bm S}$ and optimize only the remaining rotation angle and the five coefficients.
These axis-fixed runs use 40 initial evaluations and 400 BO iterations with the same 20 seed--acquisition combinations.
For $r=5$, their best CFI differs from the full-rotation value by at most $4.1\%$; for $r=10$, the difference is below $4\%$ for $N\geq10$ [Fig.~\ref{fig:axis}].
Larger reductions occur at $r=10$, $N=7$--9.
Because the constrained search is a subset of the full search, ratios above unity reflect incomplete optimization within the finite search budget, not a higher physical optimum.

The common QFI-optimal axis also moves toward the ideal finite-$N$ one-axis-twisting axis as $N$ increases.
Removing the known uniform rotation $\hat R_z(-\chi t_{\rm optm})$, we compare axes at the same preparation time using the acute angle between them.
From $N=5$ to $17$, this angle decreases from $10.48^\circ$ to $2.06^\circ$ at $r=5$, and from $54.63^\circ$ to $5.40^\circ$ at $r=10$.
The reference must be the finite-$N$, disorder-free axis: even at $N=17$ it differs from the large-$N$ asymptotic axis by $4.29^\circ$.
Together with the CFI recovery, this finite-size trend is consistent with increasing collective protection as $\chi N$ grows relative to $\sigma_\omega$; however, it neither establishes a universal crossover criterion nor determines the large-$N$ asymptotic scaling.

\begin{figure}[tb]
 \includegraphics[width=\columnwidth, bb = 0 0 244.8 201.6]{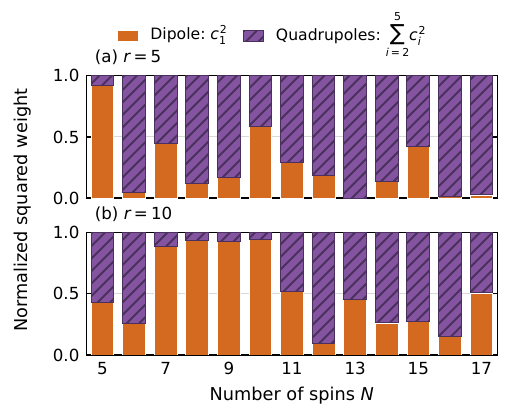}
 \caption{\label{fig:composition}Squared normalized dipole and total quadrupole coefficients of the best full-rotation BO solution for (a) $r=5$ and (b) $r=10$.
 The weights sum to one and describe the composition of the observable, not additive shares of its CFI; hatching marks the quadrupole weight.
 Near-degenerate BO solutions need not have the same composition.}
\end{figure}

Figure~\ref{fig:composition} characterizes the achieved observables through their dipole and quadrupole weights, $w_{\rm dip}=c_1^2$ and $w_{\rm quad}=\sum_{i=2}^5c_i^2$.
The balance depends on both $N$ and $r$, with substantial quadrupole contributions in the SQL-beating regime, but not quadrupole dominance at every point.
Small individual coefficients do not establish that the corresponding operators are dispensable, because CFI depends on the spectral projectors of the complete observable.
Indeed, at $r=10$ and $N=16$, a separate axis-fixed reoptimization restricted to $\hat S_y$, $\hat D_{xy}$, and $\hat Q_{xy}$ gives $\bar F_C^{\rm BO}=15.09$, compared with $\bar F_C^{\rm BO}=18.37$ for the five-component family, losing the observed SQL crossing.
Each of the 20 three-component runs uses 40 random points, 400 BO iterations, and one warm-start point projected from the five-component solution.

\emph{Discussion.---}
Local frequency disorder can degrade conventional squeezing while leaving metrological information accessible to a suitably chosen collective readout.
Our finite-size results demonstrate enhanced CFI from a restricted dipole--quadrupole family with one protocol shared across disorder realizations.
The two disorder strengths reveal a delayed SQL crossing at stronger disorder and a regime in which the common QFI-optimal axis provides an effective guide to readout optimization.
These observations motivate investigations at larger $N$, but the present size dependence is not an asymptotic power law or a quantitative extrapolation to a single macroscopic interacting ensemble.

The protocol assumes ideal spectral projective measurements of $\hat A$.
Implementing such a measurement is more demanding than measuring a few moments, and an experimental realization, finite detection resolution, and sampling noise remain to be addressed.
The QFI--CFI gap may also reflect restrictions on collective readout as well as the chosen operator family and finite optimization budget.
Within these limits, the results identify low-order nonlinear readout as a practical target for further study of metrology with inhomogeneous spin ensembles.

\begin{acknowledgments}
YC, JZ, and EY performed numerical simulations of the quantum evolution, calculated the QFI, and analyzed and interpreted the results.
EL, JZ, and EY performed Bayesian optimization of the CFI and analyzed and interpreted the results.
EY designed the study, developed the theoretical model, analyzed and interpreted the results, and wrote the original draft of the Letter.
EY thanks Janet Anders, Jes{\'u}s Rubio, and Masahito Ueda for constructive comments and discussions.
The authors used QuTiP~\cite{LAMBERT20261} for numerical simulations and the Bayesian optimization package in Ref.~\cite{bo} to optimize the CFI.
The authors acknowledge the use of ChatGPT (OpenAI) for scientific discussion, assistance in drafting numerical simulation packages and scripts, and manuscript revision.
The authors reviewed and edited the output as needed and take full responsibility for the content of this publication.
This work was supported by JST, PRESTO, Grant Number JPMJPR22BA, Japan.
The data that support the findings of this study are available from the corresponding author upon reasonable request.
\end{acknowledgments}
\bibliography{refsofnoisySSS}
\end{document}